\documentclass[twocolumn,groupedaddress,prb,showpacs]{revtex4}

\usepackage[dvips]{graphicx}
\usepackage{dcolumn}
\usepackage{bm}

\begin{document}

\bibliographystyle{apsrev}

\title{Controlled enhancement or suppression of exchange biasing using impurity $\delta$-layers}

\author{M. Ali}
\email[Email:\ ]{phyma@leeds.ac.uk}
\homepage{http://www.stoner.leeds.ac.uk}
\author{C. H. Marrows} \author{B. J. Hickey}
\affiliation{School of Physics and Astronomy, E. C. Stoner Laboratory, University of Leeds, Leeds, LS2 9JT, United Kingdom}

\date{\today}

\begin{abstract}
The effects of inserting impurity $\delta$-layers of various
elements into a Co/IrMn exchange biased bilayer, at both the
interface, and at given points within the IrMn layer a distance from
the interface, has been investigated. Depending on the chemical
species of dopant, and its position, we found that the exchange
biasing can be either strongly enhanced or suppressed. We show that
biasing is enhanced with a dusting of certain magnetic impurities,
present at either at the interface or sufficiently far away from the
Co/IrMn interface. This illustrates that the final spin structure at
the Co/IrMn interface is not only governed by interface
structure/roughness but is also mediated by local exchange or
anisotropy variations within the bulk of the IrMn.
\end{abstract}

\pacs{75.70.Cn, 75.60.-d, 75.50.Lk} \maketitle

\section{\label{intro}Introduction}

The fascination with understanding exchange bias has shown no
noticeable change, considering that 50 years have elapsed since its
discovery by Meiklejohn and Bean.\cite{Meiklejohn&bean1957}  This impetus is both practical
and fundamental,\cite{Nogues2005} since the effect both forms the
integral component of devices  such as spin-valves, magnetic tunnel
junctions, and more elaborate `spin electronic' devices, as well as
offering the opportunity to study frustration\cite{Ali2007} and the
interactions of feromagnetic and antiferromagnetic order in low
dimensions. The effect originates from the interfacial coupling of
atomic spins across a ferromagnetic (F) and antiferromagnetic (AF)
interface, the principal manifestation of which is a unidirectional
anisotropy in the F layer.\cite{Mauri1987,Tang1981,Misra2004JAP} The
main characteristic features which arise from the phenomenon are the
offset of the F magnetic hysteresis loop from zero, referred to as
the exchange bias field ($H_{\rm e}$), and its associated coercivity
enhancement ($H_{\rm c}$).

However, the precise microscopic mechanism which controls the
interfacial coupling is still a somewhat contentious topic. Large
amounts of both experimental and theoretical work\cite{Reviews1}
have highlighted the complexity of parameters which influence the
effect. It is now evident that the simple model that was first
proposed by Meiklejohn and Bean -- which assumes an ideal smooth
magnetically uncompensated surface containing a rigid spin structure
-- is inadequate in explaining the biasing. Foremost, such perfect
interfaces do not exist in reality, but moreover this model is also
unable to explain all the rich features associated with the effect,
for instance, coercivity
enhancement\cite{stiles2001,leighton2000PRL, Blamire2007PRL} and
training effects\cite{Zhang2001,Ali2003} are common to all systems
to varying degrees. It is also unable to address the asymmetrical
reversal of the magnetization in such
systems,\cite{leighton2000prl2} the AF layer thickness
dependence,\cite{sang1999,lund2002,Ali2003} and the lower than
expected experimentally obtained values for the exchange bias field.
In spite of this, it does highlight that an offset in the hysteresis
loop will only be permitted when the anisotropy of the
antiferromagnet $K_{\rm AF}$ is adequately larger than the
interlayer exchange coupling $J_{\rm AF-F}$. The importance of the
AF anisotropy was also demonstrated in an artificial exchange bias
[Co/Ru]$_{10}$/[CoPt/Ru]$_{10}$ system, where the shift in the
hysteresis loop was only shown to be present under these conditions
\cite{Steadman2002}. It has also been demonstrated that the
enhancement in the coercivity at both the onset and disappearance of
biasing are due to these terms being similar in magnitude giving
rise to a reversible magnetic component in the AF.\cite{Ali2003b}

Several theoretical models have evolved based upon the formation of domains in the AF to reduce the coupling strength.\cite{Koon1998,Schulthess1999prb,Schulthess1999apl,Nowak2002,Malozemoff1988,Misra2004JAP,Mauri1987} The most encouraging models have been those which involve  random variations in the local biasing due to defects\cite{Nowak2002,Kim2005} or roughness,\cite{Malozemoff1988} the essence of which is to dilute the spins involved, reducing the anisotropy and the exchange interaction. However, at present the interface structure is generally assumed (with very few exceptions\cite{stocks}) to be that of the bulk AF, the main reason being the extreme difficulty in experimentally ascertaining the precise structural and magnetic nature of the buried interface at the necessary atomic scale. Even for the most ideal samples it is hard to imagine that there will no re-ordering of the magnetic, crystallographic and chemical structure at the interface region. This will give rise to magnetic disorder and or spin dilution. It has been demonstrated, in an epitaxial Co/FeMn sample, how paramount the local atomic spin structure is on exchange bias.\cite{Kuch} It was shown that the atomically flat planes did not play a role, whereas the monolayer steps (atomic scale roughness) that are present at the interface  mediates the magnetic coupling across the interface. This may also resolve the quandary of why a nominally fully compensated AF surface is able to pin a ferromagnetic layer.\cite{Blamire}

However, the bulk AF spin structure also plays an important role.
Recent experiments have shown it is possible to manipulate the bias
field by ion irradiation of the
samples.\cite{Poppe2004,Mewes2000,Engel2005,Sampaio2005,Ehresmann2006}
The experiments have demonstrated that it is possible to modify the
exchange bias properties by manipulating the level of disorder
depending upon the ion dose and energy, in line with recent
theoretical models.\cite{Misra2004JAP} In the majority of these
experiments the complete system has undergone the irradiation
process including the ferromagnet. Interestingly in all cases, the
experiments have been undertaken in the presence of a external
magnetic field. This implies that system is undergoing a local
thermal treatment, where the biasing locally is being reset, hence
the necessity for an applied field. From current theoretical models
and accompanying experimental work, it is  established that there
are domains in the AF layer. However, there are still questions
regarding the formation and type. Do the domains nucleate at the
interface due to disorder, as in the domain model of
Malozemoff,\cite{Malozemoff1988} or are they more in line with
domain state model of Nowak et al.?\cite{Nowak2001}

Another class of experiments is those where spacers are introduced between the F and AF layers. The exchange bias field is essentially dependent on the relative strengths of  $K_{\rm AF}$ and $J_{\rm AF-F}$, and this has been investigated by a number of groups where spacer layers have been introduced between the AF and F layers to manipulate the strength of the coupling.\cite{GokemeijierPRL1997,ThomasJAP2000,WangJAP2002,GarciaAPL2003,ErnultJAP2003,LiJAP2003,CaiPRB2004} These studies seem to indicate that exchange bias is not necessarily a consequence of a direct exchange (nearest-neighbor) coupling mechanism. There have been contradictory reports that the exchange bias across the spacer layer is long-range in nature and decays exponentially,\cite{GokemeijierPRL1997} whilst others have reported it to be either oscillatory\cite{CaiPRB2004} or very short range in nature\cite{ThomasJAP2000}, with any long range effects ascribed to the presence of pinhole defects in the spacer.

In order to provide further insight into these questions, we report in this article on the effects of inserting a $\delta$-dusting of various elements to induce disorder at both the interface and in the bulk of the AF layer in a controlled manner. This was done by depositing a sub-monolayer of both magnetic and non-magnetic impurities in order to induce changes in the magnetic disorder on the atomic level.

\section{\label{exptl}Experimental Methods}

The Co/IrMn system was studied experimentally within a simple
spin-valve structure. A series of exchange biased spin-valve films
were deposited by dc magnetron sputtering at an argon working
pressure of 2.5 mTorr. The base pressure prior to the deposition was
of the order of $2 \times 10^{-8}$ Torr. The substrates used were
Si(100) with the native oxide layer intact, cleansed in acetone and
isopropanol. The samples were deposited  at ambient temperature, and
through masks to ensure a constant film area from sample to sample.
The system allowed 15 samples to be deposited during the same vacuum
cycle, which permitted 15 spin-valve structures Ta(75 \AA)/Co(40
\AA)/Cu(23 \AA)/Co(26 \AA)/IrMn($x$ \AA
)/$\delta$-layer/IrMn($120-x$ \AA )/Ta(50 \AA) to each specimen set
grown in indistinguishable conditions which eliminates, as far
possible, sample-to-sample variations within a run: these variations
are very small as can be seen in certain data sets later in the
paper. However, there can be more noticeable variations in these
properties from one sputtering run to the next. Hence, an important
part of our experimental methodology is to prepare an undoped
control sample in each run, to which the properties of the doped
samples can be compared. In the data presented below for the $H_{\rm
e}$ and $H_{\rm c}$ dependences on dopant layer thickness and
position in Figs. \ref{nonmaginter}-\ref{magbulk}, the 14 data
points are the doped samples from a single sputtering run, whilst
the dotted line indicates the values for these fields displayed by
the control sample.

The IrMn was deposited from a Mn target with chips of Ir attached to its surface, and energy dispersive x-ray absorption spectroscopy yielded a composition in the deposited film of $\sim$ Ir$_{25}$Mn$_{75}$. Deposition rates were determined by measuring the thickness of test films by low angle x-ray reflectometry, and were typically in the range of 2-3 \AA/s. X-ray diffraction showed that such samples are predominantly fcc with a (111) texture. We did not detect any changes in texture in a representative selection of doped samples measured by this technique, presumably since the $\delta$-layers are so thin. No post annealing steps were required, since the pinning direction was set by a 200 Oe in-plane forming field applied to the sample during the deposition of all the layers in this top spin-valve configuration.

The distance $x$ from the AF/F interface to the dopant layer was zero in some cases, but could also be an experimental variable. An IrMn layer thickness of 120 \AA\ was chosen for two reasons: the first being that it is thick enough that any fluctuations in the IrMn thickness would have a negligible effect on the exchange biasing, and the second is that it allowed the possibility of placing the impurity layer sufficiently away from the interface but still within the bulk of the layer to investigate disorder effects. Previous work has established an in-depth understanding of both the temperature and thickness dependence of the exchange bias for this Co/IrMn system\cite{Ali2003}. It has been shown that the critical thickness at which  biasing is fully established is approximately 40 \AA\ at room temperature.  For greater thicknesses, the biasing effect is found to be constant in an undoped layer. The $\delta$-layer method and the effects on giant magnetoresistance in spin-valves\cite{marrowsprb} and interlayer coupling in multilayers\cite{perezepl} has been previously described.

In comparison to the ion irradiation studies where the complete
structure undergoes irradiation, the $\delta$-dusting only generates
disorder on the atomic length scale. This also allows information on
the position dependence of the $\delta$-dusting on exchange bias.
Studies up to now have solely considered non-magnetic defects to
produce disorder. A foreign magnetic impurity will also cause both
structural and magnetic disorder through frustration for example
besides being polarized. One should be aware that even though the
particle size of  the magnetic impurities will be in the
paramagnetic regime the particle will have a Curie point dictated by
its surrounding magnetic environment through proximity exchange
effects. Hence, for particles within the IrMn layer, the Curie point
would be that of the N\'{e}el point of the IrMn, in this case
250$^{\circ}$C.

The spin-valve structure allowed the free Co layer within the spin-valve to be used as a control layer, to which the properties of the exchange coupled Co layer could be directly compared. The effect of the free layer on the pinned layer properties was minimal: orange-peel coupling fields were never more than a few Oe. It also allowed magneto-transport measurements to be performed, the resistance measurements were done using a standard four point probe dc technique. Typical (300 K) magnetoresistances of our spin-valves were $\sim 7$ per cent, whilst typical (300 K) sheet resistances were 10 $\Omega$/square. Magnetic characterization was done using a Vibrating Sample Magnetometer (VSM) and a Magneto-Optical Kerr Effect (MOKE) apparatus. All the data we shall show for $H_{\rm e}$ and $H_{\rm c}$ were acquired at room temperature.

\section{\label{results}Results and Discussion}

\begin{figure}
\begin{center}
\includegraphics[width=8cm]{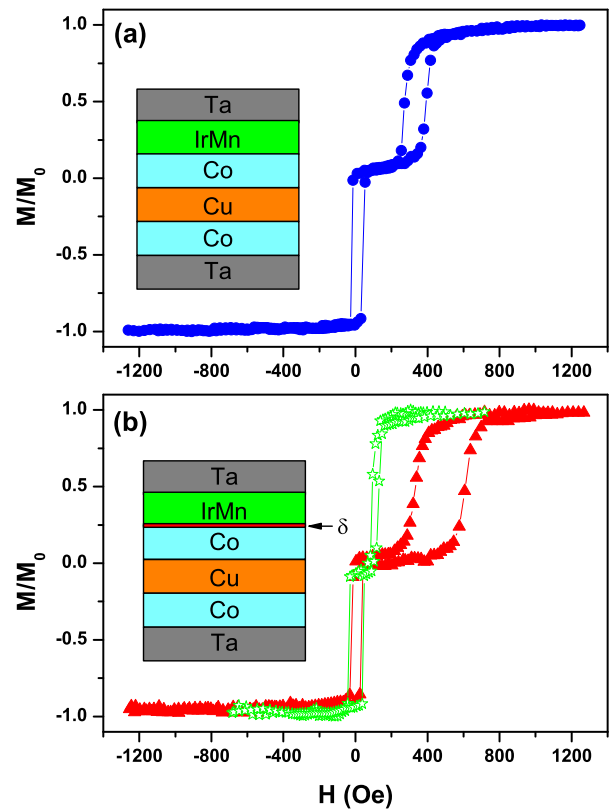}
\end{center}
\caption{(Color online) Spin-valve hysteresis loops measured by VSM: (a) undoped control sample with Co(26 \AA)/IrMn(120 \AA) exchange biased bilayer, and (b) doped samples with Co(26 \AA)/Fe(1.0 \AA)/IrMn(120 \AA) (red triangles) and Co(26 \AA)/Ta(1.0 \AA)/IrMn(120 \AA) (green stars)  pinned layers. The increase or decrease in bias bias field upon $\delta$-doping is accompanied by a commensurate change in the coercivity. \label{typicalloops}}
\end{figure}

We begin by showing in Fig. \ref{typicalloops} some hysteresis loops
that illustrate the clear spin-valve switching in our samples, as
well as the marked effects even small amounts of $\delta$-dopant can
have on the exchange bias in this system. In panel (a) we show the
typical result obtained for an undoped ``control'' sample. The
pinned and the free layer loops are easily identifiable, from which
the exchange bias field and coercivity values are straightforwardly
obtained by the usual means: $H_{\rm e}$ is the offset of the pinned
loop center from zero field, $H_{\rm c}$ is half its width. The
effects on exchange bias of a 1 \AA\ Fe or Ta $\delta$-layer at the
AF/F interface are shown in panel (b). The most striking effect is
the doubling of the exchange bias field for the introduction of the
Fe, accompanied by an enhancement in $H_{\rm c}$. Using $J_{\rm
AF-F}=H_{\rm e}M_{\rm s}t_{\rm F}$, where $J_{\rm AF-F}$ is the
interfacial exchange energy per unit area, $M_{\rm s}$ and $t_{\rm
F}$ are the magnetization and thickness of the ferromagnetic layer
respectively, values of 0.17 mJm$^{-2}$ and 0.35 mJm$^{-2}$ are
obtained for the interfacial exchange energy for the control and
Fe-doped case, consistent with much stronger exchange bonds across
the interfacial sites. Meanwhile, the introduction of Ta reduces the
interfacial exchange energy to 0.07 mJm$^{-2}$, consistent with Ta breaking
interfacial exchange bonds between Co and IrMn sites. $H_{\rm c}$ is
reduced by the introduction of Ta.

In the rest of this paper we describe in detail the effects of a selection of dopants, placed at the Co/IrMn interface, and moved away from it into the IrMn layer, on $H_{\rm e}$ and $H_{\rm c}$.

\subsection{Interfacial $\delta$-layers}

\begin{figure}
\begin{center}
\includegraphics[width=8cm]{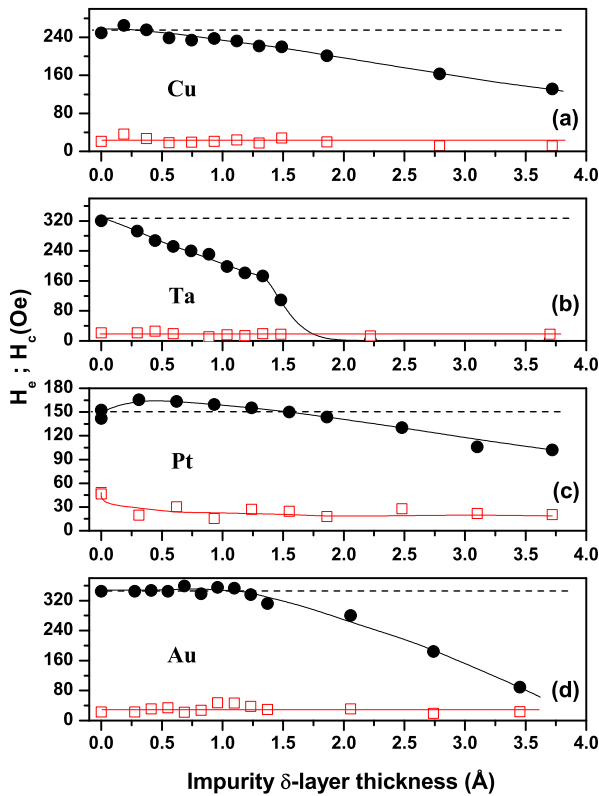}
\end{center}
\caption{(Color online) The dependence of $H_{\rm e}$ (solid circles) and $H_{\rm c}$ (open squares) on the thickness of a $\delta$-dusting of non-magnetic impurities inserted at the Co/IrMn interface. The solid lines are a guide to the eye. The dashed lines show the values obtained for undoped control samples grown in the same sputtering runs. \label{nonmaginter}}
\end{figure}

The effect of placing the non-magnetic dopants Cu, Ta, Pt and Au at
the Co/IrMn interface ($x=0$) on $H_{\rm e}$ and $H_{\rm c}$ is
presented in Fig. \ref{nonmaginter} as a function of the dusting
layer thickness. The solid lines are a guide to the eye and the
horizontal dashed lines indicate the value of $H_{\rm e}$ for the
control samples without any $\delta$-dusting. In general the
exchange bias field decreases as the dusting becomes thicker. (Here
the thickness is defined as the average equivalent thickness for the
quantity of material deposited.) It is clear that materials that
make good spacer layers for indirect exchange coupling via the RKKY
mechanism such as Cu and Au, tend to suppress $H_{\rm e}$ less
rapidly compared to materials such as Ta. For the Ta dopant there is
a monotonic decrease for thicknesses up to 1.5 \AA, before $H_{\rm
e}$ rapidly collapses to zero at that point. Interestingly this
length scale is significantly smaller than the equivalent thickness
of a monolayer (3.3 \AA), and therefore is unlikely to be a
consequence of the formation of a continuous Ta layer.

Extrapolating the curves for Cu, Pt and Au one finds that $H_{\rm
e}$ diminishes to zero at a dopant thickness of approximately 6 \AA,
8 \AA, and 4 \AA\ respectively. These thickness are greater than
that required to form a  monolayer. It could be conceived that this
is the point at which the dusting coalesces to form a continuous
layer. This is feasible, since metallic superlattice structures have
shown that it is indeed possible to obtain continuous spacer layers
of the order of 2 monolayers.\cite{Parkin1991} The small length
scales involved ($<10$ \AA), clearly suggests that exchange
interaction across the interface is very short range in nature, and
so the biasing appears to be due solely to direct exchange
interactions between spins in the F and AF layers. This is in sharp
contrast to the findings of previous work\cite{GokemeijierPRL1997}
where the exchange field was reported to exponentially decay over a
length scale of $\sim 50$ \AA. Keeping this in mind, it is even more
puzzling why a dusting of 1.5 \AA\ of Ta would destroy the biasing.
As this must be less than a monolayer, leaving large areas of direct
exchange between the Co and the IrMn layer. A possible explanation
is that a Ta atom must destroy exchange bonds involving neighboring
atomic sites as well as its own by creating an extended defect in
the electronic structure. It should be noted in systems where Ta is
placed for example next to permalloy (Ni$_{81}$Fe$_{19}$), that a chemical
reaction takes place between the two layers giving rise to a dead
layer. This has an effect of reducing the
moment\cite{Moghadam2005,Yu2005,Yang2007} of the layer.

The Pt dusting however, also exhibits an additional feature where
$H_{\rm e}$ increases by $\sim10\%$ above that for the control
sample for a $\delta$-dusting $<1.5$\AA, before then gradually
decreasing towards zero. This effect is remarkably similar to what
has been observed in perpendicular exchange bias systems, where the
addition of a Pt spacer layer is said to induce a better collinear
alignment of the Co spins out of the plane.\cite{GarciaAPL2003} This
gives rise to an increase in $H_{\rm e}$. In the present case the
spins for both layers  are confined to the easy plane of the film by
the shape anisotropy. One possibility is that the Pt is substituting
for the Ir, to form chemically ordered L1$_{0}$ phase of PtMn on a
localized basis, which itself is an AF material. PtMn possesses a
larger anisotropy, and is therefore able to orientate a larger
number of Co spins to be collinear with the unidirectional
anisotropy at the interface. This will have the effect of increasing
$H_{\rm e}$. Further increments of Pt ($>$0.4\AA) simply reduces
$H_{\rm e}$ as with the other dopants, presumably by weakening
interfacial exchange bonds.

Before moving on, we note that in all cases $H_{\rm c}$ remains
approximately constant for all dusting levels, indicating that there
is no substantial change in the AF reversible spins in the bulk or
interface, which is generally associated with any
enhancement/reduction of $H_{\rm c}$.\cite{Ali2003b} The lack of any
enhancement even at the point where the biasing vanishes (Fig.
\ref{nonmaginter}(b)), is generally interpretated the point at which
the biasing becomes reversible before vanishing with as a fuction of
AF thickness or temperature \cite{Fulcomer1972}, would imply that
the impurity is simply diluting/screening the exchange interaction
of the spins which are associated with the biasing across the F/AF
interface.

\begin{figure}
\begin{center}
\includegraphics[width=8cm]{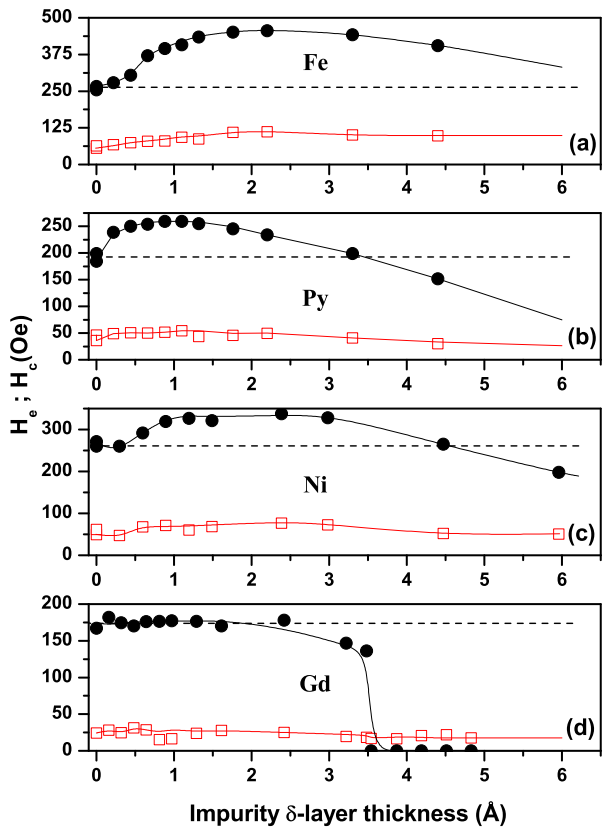}
\end{center}
\caption{(Color online) The dependence of $H_{\rm e}$ (solid circles) and $H_{\rm c}$ (open squares) on the thickness of a $\delta$-dusting of magnetic impurities inserted at the Co/IrMn interface. Solid lines are a guide to the eye. The dashed lines show the values obtained for undoped control samples grown in the same sputtering runs. The symbol Py refers to Ni$_{80}$Fe$_{20}$, the permalloy composition. \label{maginter}}
\end{figure}

The effects on $H_{\rm e}$ and $H_{\rm c}$ of placing various
magnetic dopants at the Co/IrMn interface is shown in Fig.
\ref{maginter}. Since $H_{\rm e} \sim 1/Mt$ we would expect that
increasing the total FM layer thickness by adding this material
should give a dependence where $H_{\rm e} \sim (M_{\rm Co}t_{\rm Co}
+ M_{\rm dopant}t_{\rm dopant})^{-1}$, hence decreasing the bias
field. Nevertheless, the most striking feature is the large increase
in $H_{\rm e}$ in the appropriate $\delta$-layer thickness range for
the 3$d$ metals Fe, Ni and permalloy (Py = Ni$_{80}$Fe$_{20}$). For
all these three, a broad peak in $H_{\rm e}$ approximately at
1-2\AA\ of dopant is observed. The insertion of the Fe dusting
increases $H_{\rm e}$ by some $72\%$, whereas for NiFe it is $34\%$
and for the Ni dusting the rise is $29\%$. As might be expected, the
general form of the data for the NiFe alloy falls between those for
the pure elemental Fe and Ni dopant layers. It is interesting to
note that the magnetization of the pinned layer material, Co, falls
between that of Fe and Ni. This means that Fe dopants will be
increasing the surface magnetization of the pinned layer, whilst Ni
dopants will reduce it. However both are capable of increasing the
bias field above that for a control sample. This suggests that the
increase in bias is somehow related to an inhomogeneous magnetic
interface.

We have also used a 4$f$ ferromagnet dopant, Gd, the moment of which
is known to couple antiferromagnetically to that in 3$d$ materials.
The results for Gd are shown in the bottom panel of Fig.
\ref{maginter}. Here the effect is quite different, with almost no
change in the bias field until a critical thickness of about 3.5
\AA, when $H_{\rm e}$ drops abruptly to zero. This thickness
corresponds roughly to a monolayer. Although the Gd was barely above
its bulk Curie temperature of 293 K, we should expect that it has
some ferromagnetic order as the moments will be in a strong exchange
field from the Co with which it is in intimate contact. Hence it
seems that the Gd moments do not couple to those in the IrMn which
are responsible for biasing, although why this should be so is not
clear to us at present. It should be noted that there was no
evidence of any biasing at lower temperatures of a single Gd layer.

\subsection{$\delta$-layers in the bulk}

\begin{figure}
\begin{center}
\includegraphics[width=8cm]{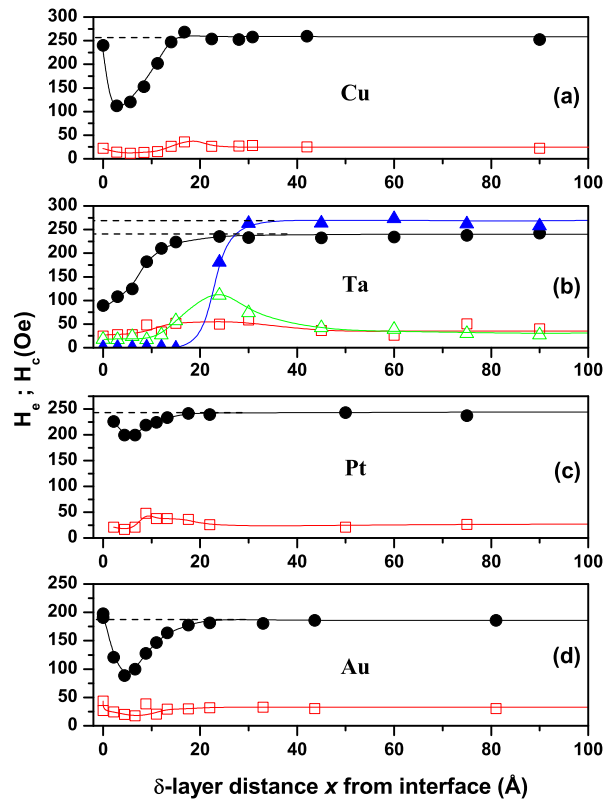}
\end{center}
\caption{The dependence of  $H_{\rm e}$(solid circles) and $H_{\rm
c}$(open squares) on the position $x$ of a 1\AA\ $\delta$-layer from
the Co/IrMn interface. Open triangles ($H_{\rm e}$) and solid
triangles ($H_{\rm c}$) in (b) represent a $\delta$-dusting of Ta of
1.5\AA. The dashed lines show the values for $H_{\rm e}$ obtained
for undoped control samples grown in the same sputtering runs.
\label{nonmagbulk}}
\end{figure}

The effects of inserting a non-magnetic $\delta$-layer of 1 \AA\
thickness into the AF layer a distance $x$ from the interface are
shown in Fig. \ref{nonmagbulk} for the same four dopant materials as
in Fig. \ref{nonmaginter}. In no case is there any increase in
$H_{\rm e}$ over the control samples. However, as the
$\delta$-dusting is moved into the IrMn for the first few \AA\ away
from the interface, $H_{\rm e}$ decreases, accompanied by a slight
increase in $H_{\rm c}$. As the $\delta$-layer is moved further
still from the interface, $H_{\rm e}$ recovers to the value shown by
the control samples once $x$ exceeds $\sim 20$ \AA. The length scale
of 20\AA\ seems to be independent of the dusting material used. The
magnitude of the dip in bias field seems again to be correlated to
the indirect exchange coupling strength of the material as a spacer
layer for RKKY coupling (Fig 2). Pt has the least effect,
followed by Au, Cu, and then Ta, which also has a detrimental effect
at the interface as shown in Fig. \ref{nonmaginter}.


This implies the $H_{\rm e}$ enhancement originates from a purely
interfacial magnetic effect, and therefore cannot be a result of
changes in the domains in the bulk of the AF of the type that is assumed in
dilution\cite{Miltenyi2000} or ion irradiation
experiments.\cite{Poppe2004,Mewes2000,Engel2005,Sampaio2005,Ehresmann2006}

We also used a slightly greater dusting of 1.5 \AA\ of Ta, which at
the interface completely suppresses the biasing, but as the dusting
is moved away from the interface $H_{\rm e}$ reappears at
approximately $x \sim 20$ \AA\, and fully recovers to that of the
control samples by $x \sim 30$ \AA. We also found that there is peak
in $H_{\rm c}$ at the onset of $H_{\rm e}$, as is usual. This
contradicts an investigation where a Au layer was moved away from
NiO/Co interface,\cite{ErnultJAP2003} where it was found that the
biasing totally disappeared as the Au layer was moved away from the
interface. The effects of the thicker Ta layer bear a striking
resemblance to the AF thickness studies that have been previously
carried out on this materials system.\cite{Ali2003b} Similar
characteristic length scales are present for the onset and
saturation of $H_{\rm e}$ along with a peak in $H_{\rm c}$ at the
onset of biasing. This suggests that the Ta layer is thick enough
here to divide the IrMn into two magnetically disconnected parts.
Only the part that is adjacent to the FM layer contributes to the
exchange bias, the other part plays no role.

\begin{figure}
\begin{center}
\includegraphics[width=8cm]{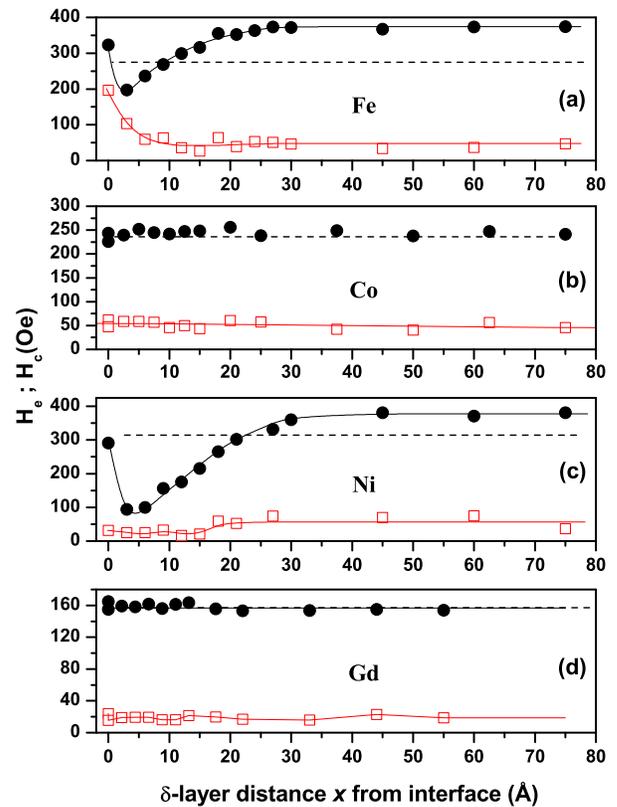}
\end{center}
\caption{The dependence of $H_{\rm e}$ (solid circles) and $H_{\rm c}$ (open squares) on the position $x$ of a magnetic $\delta$-layer of 1\AA\ thickness from the Co/IrMn interface. The dashed lines show the values obtained for undoped control samples grown in the same sputtering runs. \label{magbulk}}
\end{figure}

The effects of moving a magnetic $\delta$-layer of 1 \AA\ thickness
into the IrMn layer are shown in Fig. \ref{magbulk}: the three
elemental dopants used in the experiment reported in Fig.
\ref{maginter} appear here along with Co. The elements Gd and Co
were seen to have no significant effect on either $H_{\rm e}$ or
$H_{\rm c}$ for any value of $x$. On the other hand, there is a
clear dependence of these these two quantities on $x$ for the Ni and
Fe layers to be seen in the data. The trend is similar to that of
the non-magnetic $\delta$-layers for small values of $x$, where
there is a dip in $H_{\rm e}$ at approximately $x=5$ \AA. However,
as the layer is moved further from the interface not only does the
biasing recover and saturate by 30 \AA, the magnitude also increases
in comparison to the control samples by 20\% for Ni and 34\% for the
Fe. In general no obvious trend with the position of the
$\delta$-layer is evident in $H_{\rm c}$. There is enhancement in
$H_{\rm c}$ when the Fe is present at the interface, but this falls
rapidly back to the control sample level once beyond 5 \AA. This is
also evident in Fig. \ref{maginter}, where a slight increase in
$H_{\rm c}$ is observed.

As with the non-magnetic elements, the dip in $H_{\rm e}$ is
attributed to the dilution of the interfacial magnetic moment and
anisotropy for the Fe and Ni elements: given that these are magnetic
elements we should not expect a significant depression in the local
exchange interaction strength. One can speculate that the $\delta$-layer is neutralizing the uncompensated
moments associated with the biasing. It may be
significant that whilst FeMn,\cite{ishikawa1974}
CoMn,\cite{cable1994} NiMn,\cite{kren1968} and
GdMn\cite{goncharenko2005} all have antiferromagnetic phases, only
FeMn and NiMn show a significant exchange bias at room
temperature.\cite{lin} Exchange bias from antiferromagnetic CoMn is
generally either non-existent or weak.\cite{huangjmmm2000} We are
unaware of any reports of attempts to observe an exchange bias using
a GdMn-based AF layers. At the interface the Fe and
Ni simply couple F with the Co layer, whereas
immediately within the IrMn layer they couple AF with the uncompensated spins in the vicinity of the interface,  in this manner
effectively reducing the net interfacial magnetization.
Elements such as Au or Cu reduce the biasing because they possibly form
the classical spin glass phases of CuMn\cite{Ali2007} and AuMn. At room temperature the spin glass
would behave as paramagnetic entity and similarly reduce the net interfacial magnetization.
For these reasons one might obtain a dip in $H_{\rm e}$ as function of position.
Away from the interface the Ni and Fe create an additional AF system (FeMn/NiMn)
within the IrMn which enhances the biasing. What is intriguing is the lack of any
effect of the Co or Gd on the $\delta$-layer position. One can only
infer that the Co and the Gd atoms are easily accommodated into the
magnetic structure of the IrMn layer for the dusting levels employed
and therefore have a negligible effect on the local anisotropy.

The results of Fe and Ni seem to suggest that not only is the
interfacial anisotropy paramount for exchange biasing (the dip), but
the final magnetic state is also influenced by the bulk magnetic
state of the AF layer due to the enhancement in $H_{\rm e}$ beyond
30 \AA. These results seem to be in agreement with the diluted
domain state models and the ion irradiation experiments.

\section{\label{conc}Conclusion}

We have shown that magnetic disorder is a key ingredient in
understanding the exchange bias phenomenon by studying the effects
of inserting impurity $\delta$-layers of various elements at both
the Co/IrMn interface and at given points within the IrMn layer
itself. The experiments have shown the importance of disorder in the
vicinity of the interface and throughout the bulk of the AF layer,
and is consistent with the domain state model. By using both
magnetic and non magnetic $\delta$-layers, it is possible to
conclude that it is the magnetic disorder which seems to dominate
and control the exchange bias effect.  Any effect which is able to
generate magnetic disorder will therefore influence the exchange
bias. In general non-magnetic elements were found to reduce the
exchange coupling, the exception being Pt where larger anisotropies
are induced. On the other hand, when placed correctly, the magnetic
elements induced a stronger exchange bias due to the increase in
magnetic disorder by inducing stronger exchange bonds or anisotropy
at the doped atomic sites. Overall, we have observed a rich variety
of behavior that we hope will provide a spur to the development of
theories that treat disorder in exchange bias systems. Also, these
results demonstrate a means of tailoring and improving the magnitude
of exchange anisotropy in device applications.

\begin{acknowledgments}
This work was supported by the EPSRC.
\end{acknowledgments}

\end{document}